# Sub-Wavelength Plasmon Polaritons Channeling of Whispering Gallery Modes of Fluorescent Silica Microresonator

*Sunny Tiwari,[1,†,*] Chetna Taneja,[1,†,*] G. V. Pavan Kumar[1,2,*]*

[1]Department of Physics, Indian Institute of Science Education and Research, Pune, India-411008.

[2]Center for Energy Science, Indian Institute of Science Education and Research, Pune, India-411008.

[*]Corresponding authors: sunny.tiwari@students.iiserpune.ac.in

chetna.taneja@students.iiserpune.ac.in and pavan@iiserpune.ac.in

[†]These authors contributed equally to this paper.

**Abstract:**

Herein, we report sub-wavelength propagation and directional out-coupling of whispering gallery modes (WGMs) of a fluorescent silica microsphere resonator mediated via plasmon polaritons in a single crystalline silver nanowire. Fluorescent spectral signatures of emission from the nanowire ends confirm efficient guiding of WGMs via nanowire plasmons. By employing Fourier plane optical microscopy, we reveal and quantify the directional fluorescence emission of WGMs from the ends of the nanowire. Given that the discussed geometry is self-assembled from a solution-phase, our results will find relevance in WGM-based soft-photonic platforms including miniaturized optical resonators and couplers.

## 1. Introduction:

Combining plasmonic and dielectric photonic structures can offer interesting opportunities in hybrid nanophotonics [1-3]. This paper shows how plasmonic nanowires combined with dielectric microsphere can be harnessed to realize a directional source of whispering gallery modes.

One of the most essential entities in nanophotonics devices is to collect and channel light at scales comparable to the wavelength of light [4]. These requirements cannot be achieved by traditional optical elements such as lenses, mirrors or gratings because of the diffraction limit. Thus, we look for alternative sources to confine and direct light at nano/microscale. Metallic nanostructures support surface plasmons, which lead to field enhancement at the metal/dielectric interface and between two metallic structures. These sub-wavelength hot-spots of high local electric fields have applications such as enhancing molecular emission [5, 6], cavity electrodynamics [7] and bio-imaging beyond diffraction limit [8]. Although, metallic nanostructures can focus and channel light at sub-wavelength scale, they suffer huge losses at optical frequencies.

Certain dielectric cavities on the other hand, facilitate high quality factor whispering gallery modes (WGMs), which have been utilized for various applications such as lasing [9], bio-chemical sensors [10], non-linear photonics [11, 12] and curved light beams [13]. Among all the dielectric structures supporting WGMs like micro-rings, toroids, micro-disks, dielectric microspheres [14-16] have gained a lot of interest due to the ease in the fabrication. A large number of the studies have been focused on examining spectral and polarization signatures of these modes [14, 17, 18]. Wavevectors of WGMs emission, which is crucial for device related application is still relatively less explored [19-21]. WGMs emission from an isolated dielectric microsphere are isotropic in nature because of the symmetry of dielectric cavities. Isotropic emission results in the low collection efficiency and inefficient coupling at sub-wavelength scales. In the past, dielectric microspheres have been coupled to dielectric waveguides for evanescent excitation of WGMs [22, 23]. However, diffraction limited waveguides cannot support sub-wavelength propagation of WGMs and the emission from these coupled structures is non-directional in nature.

Recently, we have studied a hybrid metallo-dielectric structure for the remote excitation of WGMs of dielectric structures using metallic silver nanowire [24]. It was shown that the WGMs emission from dielectric structure coupled to metallic nanowire is direction in nature. Although the directionality was achieved in remote excitation configuration, the subwavelength propagation of whispering gallery modes through nanowire plasmons was not explored.

In this paper, we utilize a hybrid, self-assembled silver nanowire- $SiO_2$ microsphere (AgNW-μS) junction for sub-wavelength propagation of microsphere WGMs while simultaneously tuning the directionality. We study the spectral and wavevectors signatures of WGMs supported by molecules coated dielectric microsphere coupled to a plasmonic nanowire. Light coupled at one end of the metallic nanowire out-couple from the other end or at any discontinuity along the nanowire [25, 26]. Coupling chemically prepared single crystalline silver nanowire [27] to a dielectric microsphere opens up the possibility of waveguiding WGMs at sub-wavelength scale and providing directionality to WGMs wavevectors. Using the near field coupling of WGMs emission from the microsphere with nanowire plasmons, we procure directionality in the emission of WGMs which is otherwise not possible from an isolated microsphere.

**2. Experimental schematic of the study**

The schematic of the experimental configuration is shown in Fig. 1 (a). Molecules coated dielectric microsphere (μS) **is** coupled to chemically prepared silver nanowire (AgNW) using self-assembly technique [28, 24]. Upon excitation of the junction, the emission from the distal end of the nanowire **is** studied both in terms of spectral and wavevectors signatures and compared with the emission from isolated microsphere. Figure 1b (i) shows the optical bright field image of dye coated isolated microsphere of ~3 µm diameter placed on a glass substrate and 1b (ii) shows the inelastic scattering image of emission from the microsphere. Nile blue molecules coated microsphere of diameter ~3 µm coupled to a silver nanowire of thickness ~350 nm is shown in Fig. 1c (i). Upon excitation of the junction, emission at the junction and at the distal ends of silver nanowire can be seen in the elastic scattering image of the AgNW-µS system (Fig. 1c (ii)). The distal end of silver nanowire is spatially filtered as shown in Fig. 1c (iii) and the inelastic image of the spatially filtered distal end is shown in Fig. 1c (vi).

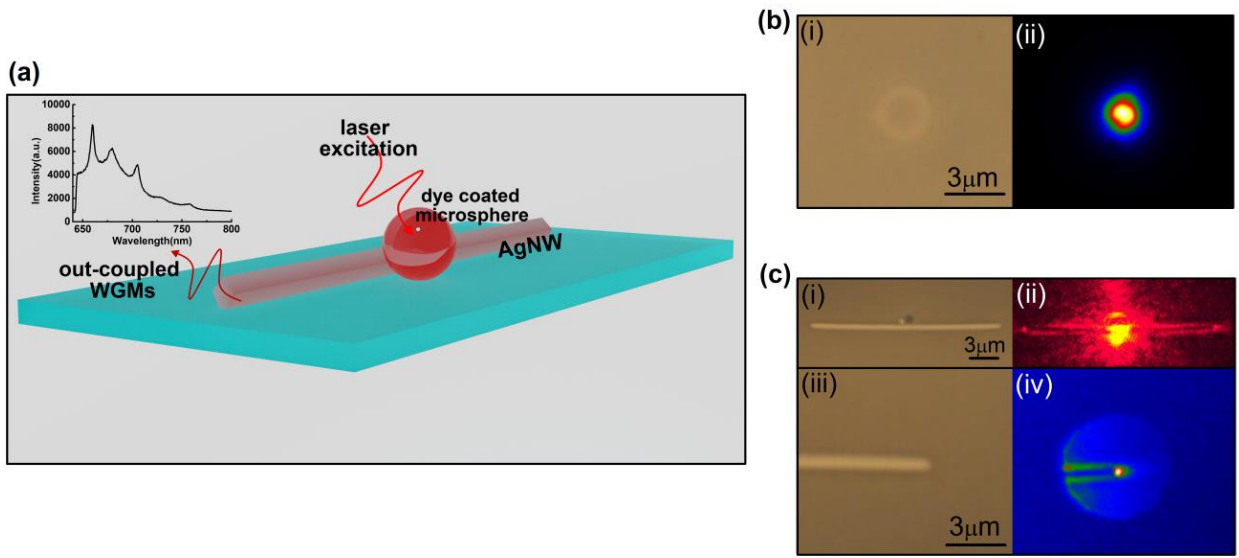

*Fig. 1. (a) Schematic of the experiment configuration. A dye coated microsphere is assembled near a silver nanowire using self-assembly. The nanowire-microsphere junction is excited with a 633 nm laser. The emission from the molecules excites the whispering gallery modes (WGMs) of the sphere which gets coupled to nanowire plasmons. At the distal end of nanowire, plasmons get out-couple as WGMs of the microsphere in a directional manner. Comparison between the excitation of WGMs of (b) an isolated dielectric microsphere placed on glass substrate (c) microsphere*

*coupled with silver nanowire. (b) (i) Bright field (b) (ii) inelastic scattering images of an isolated dielectric microsphere placed on a glass substrate. (c) (i) Bright field image of AgNW-µS junction. (c) (ii) Elastic scattering image of the system upon excitation of AgNW-µS junction. (c) (iii) Zoomed-in bright field image of one end of silver nanowire shown in (c) (i). (c) (iv) Inelastic scattering at the spatially filtered end of the nanowire shown in (c) (iii).*

## 3. Results and discussion

For the experimental studies, AgNW-µS junction is excited with 633 nm laser wavelength polarized along the junction using 100x, 1.40 objective lens and the backscattered light is collected using the same objective lens. For the complete experiment setup refer to our previous works [24, 29].

### 3.1 Spectral analysis of fluorescence emission coupled to WGMs

The microsphere acts as a scatterer to couple light into the nanowire plasmons. These propagating plasmons travel along the length of silver nanowire and out-couple as free photons at both the ends of nanowire which can be seen from the elastic scattering image of AgNW-µS system in the Fig. 1c (ii). Since Nile blue molecules coated on the microsphere is fluorescent around excitation laser, the fluorescence of the molecules also gets excited at the junction which couples to the WGMs of the microsphere. For the spectral study of the inelastic emission, junction and distal end are spatially filtered using a pinhole and the collected light is projected to the spectrometer after rejecting the elastically scattered light. The black solid and red dashed curves are spectra collected from the junction and nanowire end respectively. Both the spectra are collected for different exposure time as the intensity of the emission is different at both spots.

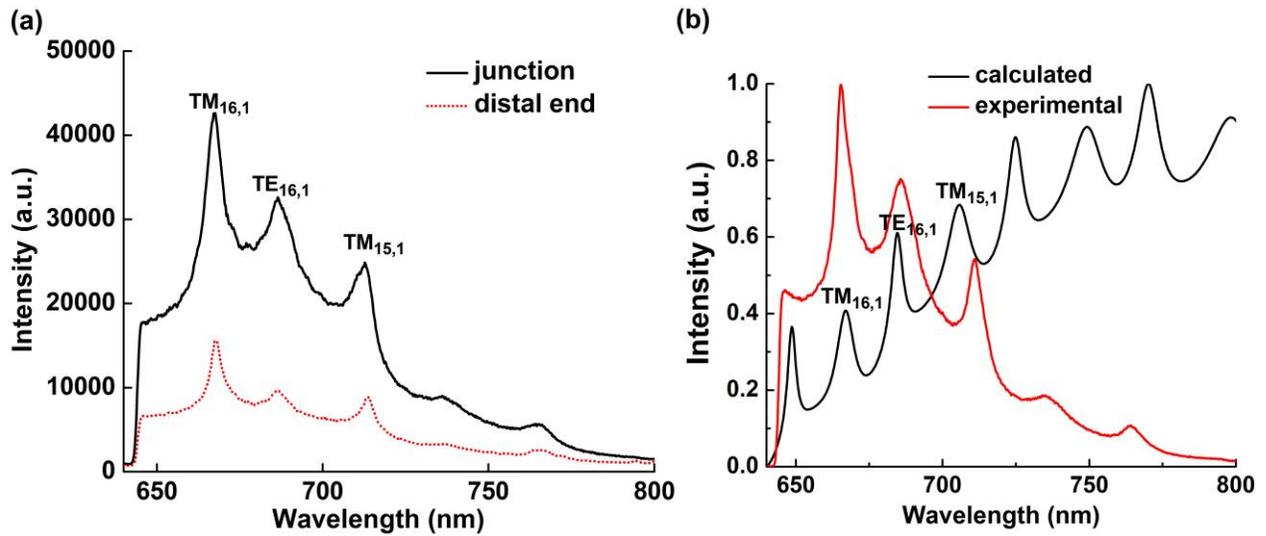

*Fig. 2. Spectral signatures of whispering gallery modes. (a) WGMs of a microsphere coupled to nanowire. Black and red curves show the spectral signatures of the emission collected from the AgNW-µS junction and nanowire end respectively. (b) WGMs of an isolated microsphere. Red curve shows the experimentally obtained WGMs of an isolated dye coated microsphere placed on a glass substrate. Black curve shows the calculated WGMs of an isolated microsphere in a medium of refractive index 1.*

Spectra collected at the junction, shows WGMs of the microsphere coupled to fluorescence of the Nile blue molecules. WGMs are labelled using Mie theory as transverse electric (TE) and transverse magnetic (TM) based on polarization of the mode. Modes (TE/TM $_{l,n}$) are identified using three quantum numbers n, l and m. n, radial quantum number quantifies the intensity maxima along the radius of the microsphere. l, orbital quantum number represents the number of halves of the intensity maxima along the circumference of the microsphere. Projection of the orbital quantum number on the quantization axis is called azimuthal quantum number m. Modes of WGMs are more intense around the fluorescence maxima of Nile blue molecules.

It is interesting to note that the spectrum collected from one of the distal ends of the silver nanowire also show sharp resonances where the peak positions represents the TE and TM polarized WGMs of the microsphere. This suggests that WGMs of the microsphere excited at the junction couple to the nanowire plasmons and out-couple at the nanowire end. For comparison, we show the WGMs of an isolated microsphere placed on a glass substrate (red curve, figure 2b) and calculated WGMs of an isolated microsphere in vacuum using the Mie scattering method for a refractive index of 1.45 (black curve, figure 2b) [30]. The slight difference in the peak positions of the isolated microsphere and microsphere placed on the glass substrate is because of the interaction of the microsphere with the substrate which causes a red shift in the spectrum [31].

## 3.2. Directionality of remotely collected WGMs

Spectral analysis confirms that WGMs of the microsphere coupled to a silver nanowire can be collected remotely at the nanowire end. The next step is to study the directionality of the WGMs wavevectors at the distal end of nanowire. For this, we perform Fourier plane imaging, which essentially maps the back focal plane of an objective lens and quantifies the directionality of emission wavevectors in terms of radial and azimuthal angles [32]. The Fourier plane image of the inelastic emission from the spatially filtered distal end of the nanowire is shown in the Fig. 3c. It is evident that out-coupled WGMs wavevectors are directional in nature. For the qualitative analysis, we compare directionality of the WGMs wavevectors out-coupled at the nanowire end with the WGMs wavevectors of an isolated dielectric microsphere of same size placed on a glass substrate.

Figure 3a (i) and 3a (ii) show the bright field and inelastic scattering image of the spatially filtered distal end of nanowire coupled to a microsphere upon excitation of the junction. Figure 3b (i) and 3b (ii) show the bright field and inelastic scattering image of an excited isolated microsphere placed on a glass substrate. The Fourier plane images shown in Fig. 3 (c) and 3 (d) correspond to the emission shown in Fig. 3a (ii) and 3b (ii) respectively. Intensity cross-cut along white dotted line, in Fig. 3 (c), shows radial angle distribution of the WGMs emission at the nanowire end and reveals a sharp emission near the critical angle of glass-air interface. For the azimuthal spreading of the emission from the nanowire end, polar plot of intensity along the circumference of the circle at critical angle is plotted in Fig. 3 (f). The emission covers very narrow range of azimuthal angles.

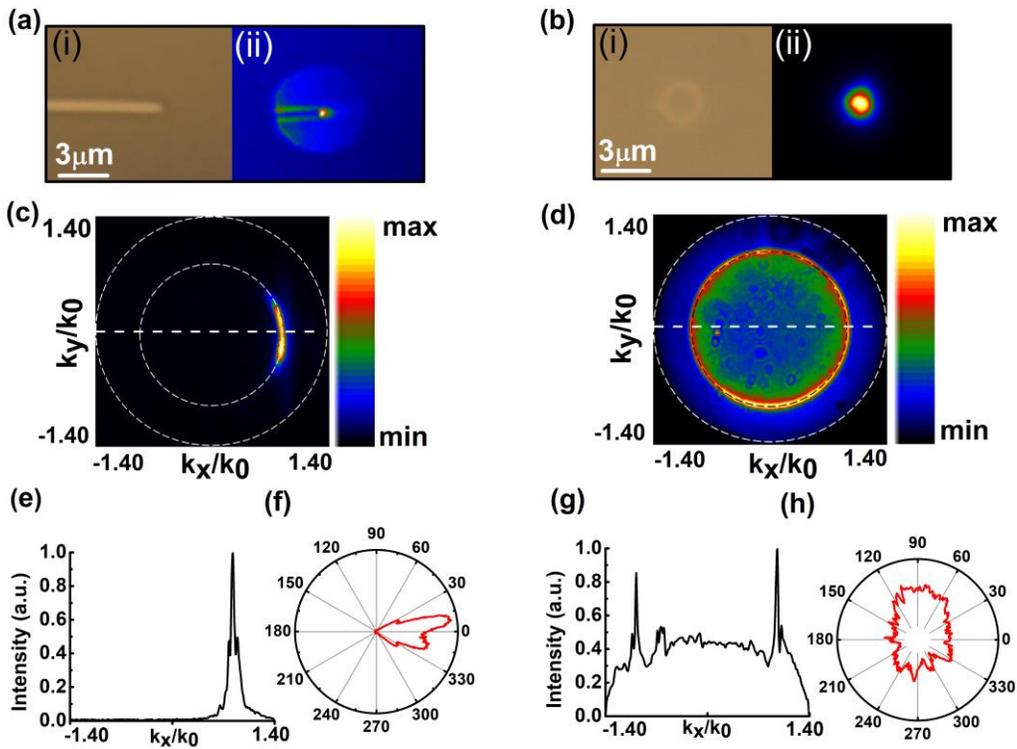

*Fig. 3. Whispering gallery modes wavevector comparison of emission from nanowire end with the emission from isolated microsphere. (a) (i) Bright field (ii) inelastic scattering images of spatially filtered nanowire end. (c) Fourier plane image of the emission collected from distal end of*

*nanowire shown in (a) (ii). (e) Intensity cross-cut along the white dotted line shown in (c) representing out-coupled wavevectors near the glass-air critical angle. (f) Azimuthal intensity distribution along the inner circle in (c) at a constant radial angle. (b) (i) Bright field (ii) inelastic scattering images of a molecule coated microsphere placed on a glass substrate. (d) Fourier plane image of inelastic emission from microsphere. (g) Intensity cross-cuts along the white dotted line shown in (d), (f) azimuthal intensity distribution along the inner circle in (d) at a constant radial angle.*

On the other hand, the Fourier plane image of emission from an isolated microsphere shows isotropic emission covering large azimuthal angles. Quantitatively, a cross-cut along white dotted line shows the emission is maximum near the glass-air interface critical angle (Fig. 3 (g)) but is also comparable at other angles. In terms of azimuthal angles, the emission covers almost all the angles as shown in Fig. 3 (h). Overall, the spreading of WGMs wavevectors at the nanowire end is very narrow in comparison to the emission from isolated microsphere, which makes the AgNW-µS hybrid system a directional source.

### 3.3. Selectively probing the directional WGMs at nanowire ends

Inelastic scattering image from the AgNW-µS system (Fig. 4a (ii)) shows emission from both ends of nanowire, which can be used to get directional WGMs in different directions. For this, we selectively probe the directionality of emission from both the ends. Ends of the nanowire are marked as end 1 and end 2, visible in the optical bright field image of AgNW-µS system in Fig. 4a (i). Selectively, each end of the nanowire is spatially filtered using a pinhole and projected to a spectrometer for spectral analysis. Black (solid) and red (dashed) curves in Fig. 4 (b) show the spectra collected from end 2 and end 1 respectively after rejecting the elastically scattered light. The relatively higher intensity of end 1 in comparison to end 2 can be attributed to different distance of both ends from the junction. Spectra confirms out-coupling of WGMs of the microsphere coupled to silver nanowire through both the ends. The Fourier plane images of emission from end 1 and end 2 are shown in Fig. 4 (c) and 4 (d) respectively. The emission covers a small range of both radial and azimuthal angles and can be tuned in either of the directions along the nanowire depending on the nanowire end selected.

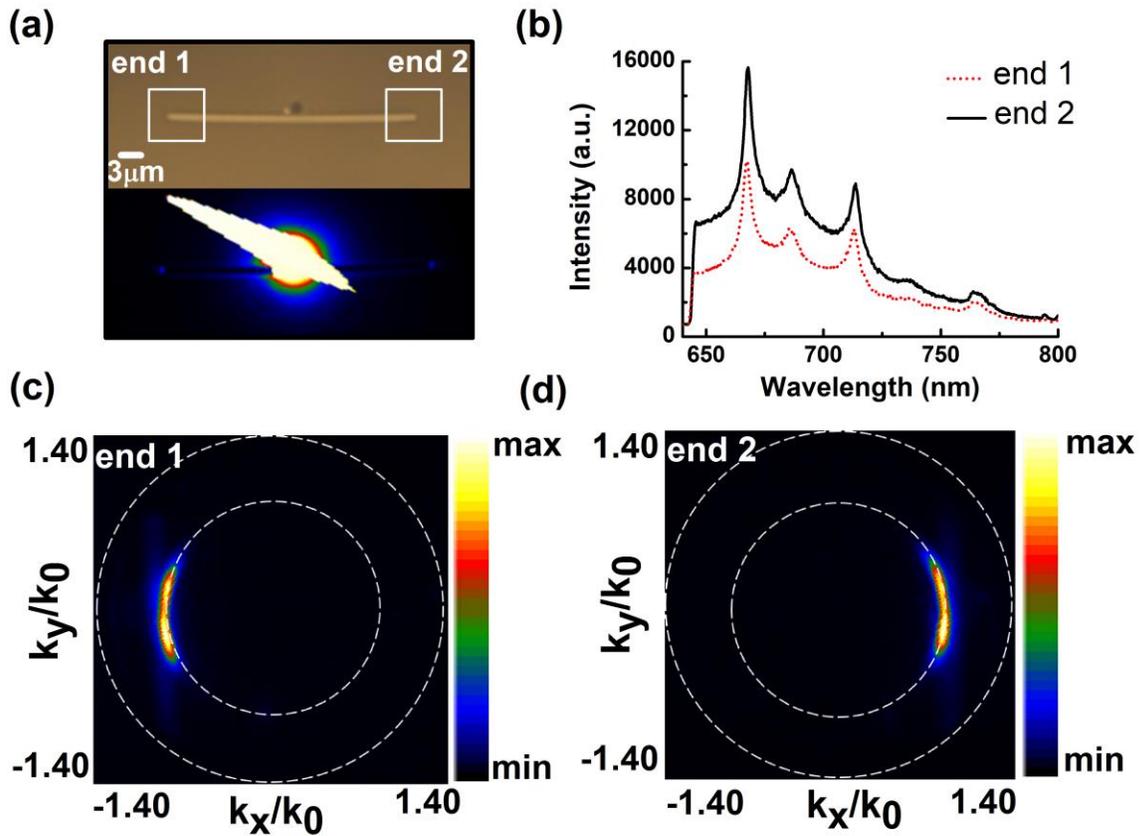

*Fig. 4. Fourier plane imaging of WGMs wavevectors from both the ends of nanowire. (a) Bright field and inelastic scattering images showing the WGMs emission at the excitation point as well as at both ends of silver nanowire upon excitation of the junction. (b) Spectral signatures of the emission at end 1 and end 2 of the silver nanowire as labelled in (a) with white square boxes. Red (dashed) and black (solid) curves show the spectra collected individually from spatially filtered end 1 and end 2 respectively. Fourier plane images of the emission from (c) end 1 and (d) end 2.*

### 3.4. Coupling of WGMs into nanowire mode

To understand the coupling of these WGMs into silver nanowire, we perform plasmon leakage radiation microscopy [33] on the emission from AgNW-µS system without any spatial filtering. Fig. 5 (a) shows the optical bright field image of the AgNW-µS system in transmission mode. The AgNW-µS is excited with a 633 nm laser polarized along the junction using 100x, 0.95 objective lens and the forward scattered light is collected using 100x, 1.49 oil immersion objective lens. Elastic scattering image of the AgNW-µS system in Fig. 5 (b) shows the emission at the junction and at the distal end of nanowire. Along with the emission, leaky mode of the nanowire at excitation wavelength in bright field image can also be seen along the length of the nanowire [34]. Fig. 5 (c) shows the spectral signatures of the emission from the AgNW-µS system after rejecting the excitation wavelength. The emission is indeed WGMs of the microsphere coupled to the fluorescence of Nile blue molecules. This inelastic emission is sent to EMCCD for leakage radiation microscopy without any spatial filtering whose Fourier plane image is shown in Fig. 5 (d). Fourier plane imaging shows that the majority of the emission is near the air-glass critical angle. In addition, two straight lines along $k_y/k_o$ at constant values of $k_x/k_o$ are also prominent. These two straight lines represent leaky modes of nanowire. This confirms the coupling of whispering gallery modes of microsphere to nanowire plasmons.

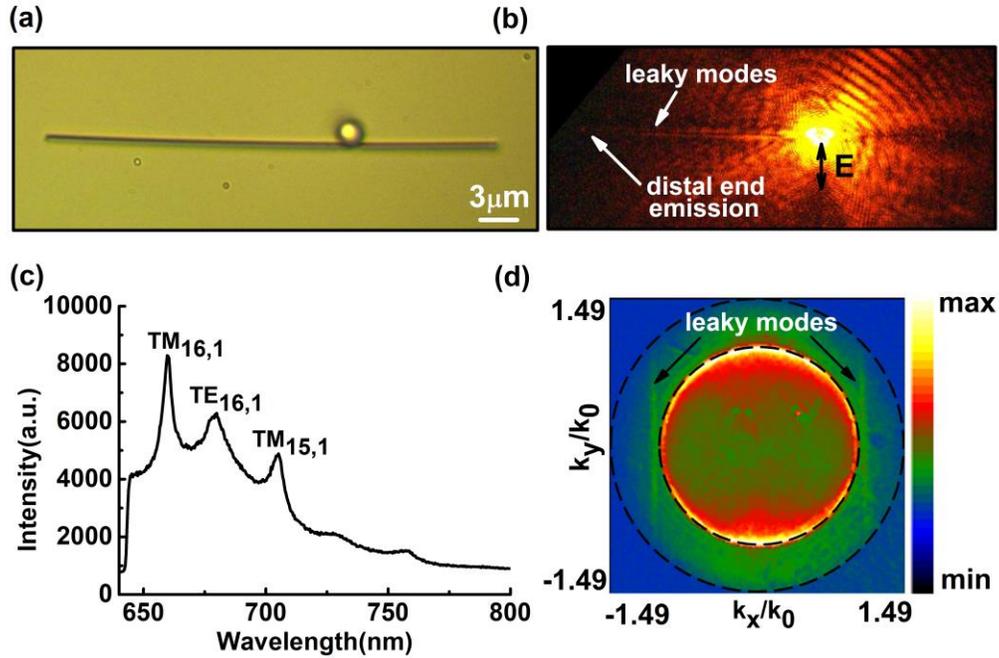

*Fig. 5. Spectral and wavevector analysis of WGMs of dielectric microsphere coupled to nanowire plasmons. (a) Bright-field image of silver nanowire-dielectric microsphere (AgNW-µS) system in transmission mode where microsphere of diameter ~3 µm is coupled to a ~350 nm thick nanowire. (b) Elastic scattering image of the system upon exciting the AgNW-µS junction with 633 nm wavelength laser. (c) Spectral signatures of emission showing WGMs of microsphere riding over the fluorescence. (d) Fourier plane image of emission from the full system of nanowire coupled to the microsphere. Two leaky modes in the Fourier plane image represent the coupling of WGMs to the nanowire plasmons propagating in opposite directions.*

## 4. Conclusion

In summary, we show the sub-wavelength propagation and remote collection of WGMs mediated by silver nanowire plasmons. Fourier plane imaging is performed to study the directionality of wavevectors of emission from nanowire ends. Leakage radiation microscopy is also performed to understand the mechanism behind the directional emission of WGMs from nanowire ends. The WGMs of microsphere couple to the nanowire as propagating modes and out-couples from the nanowire end. The emission is directional in nature and spans a very narrow range of radial and azimuthal angles making the system a directional hybrid antenna.

**Funding**

This work was funded by Air Force Research Laboratory grant, DST Energy Science (SR/NM/TP-13/2016) and DST Swarnajayanti fellowship (DST/SJF/PSA-02/2017-18).

**Acknowledgment**

Authors thank Diptabrata Paul for preparing the nanowires and Rafeeque (Science and media center, IISER Pune) for drawing the schematic. Authors also thank Dr. Deepak K Sharma and Vandana Sharma for fruitful discussions. Chetna Taneja acknowledges INSPIRE fellowship for the funding.